\documentclass[5p,superscriptaddress]{elsarticle}

\usepackage{amsmath,amssymb}
\usepackage[utf8x]{inputenc}
\usepackage{ucs}

\usepackage{amsfonts}
\usepackage{dcolumn}
\usepackage{bm}
\usepackage[usenames]{color}
\usepackage{multirow}
\usepackage{graphicx}
\usepackage{hyperref}
\usepackage{subfig}
\usepackage{booktabs}
\usepackage{xcolor}
\usepackage[normalem]{ulem}

\usepackage{float} 
\usepackage{wrapfig} 

\usepackage{upgreek} 
\usepackage{cancel} 
\usepackage{mathdots} 
\usepackage{mathrsfs} 

\begin{document}

\title{Monotone measures of statistical complexity}

\author[fr,cft]{{\L}ukasz Rudnicki}
\author[g1,g2]{Irene V. Toranzo}
\author[g1,g3]{ Pablo Sánchez-Moreno}
\author[g1,g2]{Jes\'us S. Dehesa}
\address[fr]{Institute for Physics, University of Freiburg, Rheinstra{\ss}e 10, D-79104 Freiburg, Germany}
\address[cft]{Center for Theoretical Physics, Polish Academy of Sciences, Aleja Lotnik{\'o}w 32/46, PL-02-668 Warsaw, Poland}
\address[g1]{Instituto {\em Carlos I} de F\'isica Te\'orica y Computacional, Universidad de Granada, 18071-Granada, Spain}
\address[g2]{Departamento de F\'isica At\'omica, Molecular y Nuclear, Universidad de Granada, 18071-Granada, Spain}
\address[g3]{Departamento de Matemática Aplicada, Universidad de Granada, 18071-Granada, Spain}

\date{\today}

\begin{abstract}
 We introduce and discuss the notion of monotonicity for the complexity measures of general probability distributions, patterned after the resource theory of quantum entanglement. Then, we explore whether this property is satisfied by the three main intrinsic measures of complexity (Crámer-Rao, Fisher-Shannon, LMC) and some of their generalizations. 
\end{abstract}



\maketitle

\section{Introduction}

On which grounds shall one build a sound description of complexity? Despite the great efforts done in many areas of science ranging from atomic, molecular an nuclear physics up to the adaptive complex systems and ultimately the living beings  \cite{lloyd,gell-mann,badii,shiner,yamano_jmp04,yamano_pa04,catalan_pre02,holland,sen_2012,sanchez-moreno,tan-vedral}, this question does not have a definitive answer supported by a systematic treatment. 
Intuitively, the complexity of a finite many-particle system is a measure of the internal order/disorder of the system in question, which must be closely connected with the notion of information and its main quantifier, namely the information entropy.  Interpreting the second law of thermodynamics, which indicates an always increasing entropy, one can vaguely explain the fact that information entropy is maximal for a completely disordered system. 

The complexity, however, behaves in a completely different manner. 
A completely ordered or completely regular system (e.g., a perfect crystal) is obviously \textit{non-complex}, but also the structure of a completely disordered or absolutely random system (e.g, an ideal gas) enjoys a very simple description. We say that these two extremal cases have no complexity, or rather an extremely low, minimum complexity. A complexity quantifier applicable to physical systems with different degrees of order/disorder, 
the great majority of which lie down somewhere between the two extremes, shall take the above observation into account. In other words, a proper \textit{measure of complexity} (with a suitably chosen, scenario-adapted notion of non-complexity) shall assume the minimal value for a non-complex input.

But the theory of quantum entanglement \cite{vidal} together with recent developments made for quantum coherence \cite{BCP14}, point out that being discriminative with respect to separable or incoherent states is not enough to be a faithful measure of the resource. The measure of entanglement (coherence) needs to be a monotone which does not increase under LOCC (incoherent) operations. This means, that one cannot generate entanglement or quantum coherence by performing only "classical" or "local" operations.

The aim of this contribution is to transfer the intuition related to quantification of entanglement and coherence, to the field of statistical complexity. If one wishes to quantify the complexity of a certain probability distribution, one shall 
take into account the fact that  the distribution  at hand can be smeared in a way that it becomes 
closer to the description of the non-complex system. A good measure of complexity shall then assign a lower value to such new, transformed distribution. Note that we do not mean here any possible smearing, but only transformations which lead to less complex distributions.  

More formally, we say that a complexity measure $C[\rho]$ defined for a single-particle probability density $\rho(x)$ is a monotone in the following sense:
\begin{itemize}
\item[(i)] There exists a family $\Xi$ of densities with minimal complexity, so that  if $\rho\in\Xi$, then $C[\rho] \le C[\rho']$ for any other density $\rho'$.

\item[(ii)] There exists a class of operations $\mathcal{G}$ that preserve $\Xi$, i.e. if $\rho\in\Xi$, then $\mathcal{G}[\rho]\in\Xi$.

\item[(iii)] The complexity measure $C$ is monotonic with respect to all operations from the class $\mathcal{G}$, what means that $C[\mathcal{G}[\rho]]\le C[\rho]$ for any density $\rho$.
\end{itemize}
These three properties capture the idea of monotonicity as described above. Following the theory of quantum entanglement we shall now postulate that the proper measure of statistical complexity satisfies the monotonicity requirement. Note that the "axioms" (i--iii) provide only a general framework for studying measures of statistical complexity. Working with particular scenarios one always first needs to specify the notion of being non-complex and think of the operations which potentially only decrease the complexity. It shall turn out that various measures are monotones with respect to different couples $(\Xi,\mathcal{G})$, what physically implies that they are able to properly describe distinct emanations of complexity.

In the last few years various measures of the complexity of a finite many-particle system have been suggested in terms of two spreading measures (e.g., variance, Shannon entropy, Fisher information, disequilibrium) of a single-particle probability density. The most important examples are the complexity measures of Crámer-Rao \cite{dembo,dehesa_1,antolin_ijqc09}, Fisher-Shannon \cite{VignFS,angulo_pla08,romera_1} and LMC (López-ruiz, Mancini and Calvet)\cite{catalan_pre02}. In Section 2 we briefly review the construction of these measures, while in Section 3 we use them to explain the idea behind the postulated monotonicity. In particular, we show that both Crámer-Rao and  Fisher-Shannon measures are monotones with respect to a convolution with any Gaussian probability distribution, while the discrete LMC complexity measures are monotone with respect to all stochastic operations preserving the class containing the uniform distribution and the Kronecker delta distributions. Note that while the former case seems to be complete, the monotonicity of LMC complexity in the continuous scenario, even though very likely to occur, is left as an open question.




\section{Basic complexity measures}

Let us consider a general one-dimensional random variable $X$ characterized by the continuous probability distribution $\rho(x)$,
$x \in \Lambda \subseteq \mathbb{R}$, which is assumed to be normalized so that $\int_{\Lambda} \rho(x) dx  =  1$.
%
The information theory provides various spreading measures of the distribution beyond the familiar variance $V[\rho]$, such as the well-known Shannon entropy \cite{shannon_49}
\begin{equation}
     S[\rho] =-\int_{\Lambda} \rho(x) \ln \rho(x) dx,
  \end{equation}
the Rényi entropy of order $\lambda$  \cite{renyi_70} given by
   \begin{equation}
       R_\lambda[\rho] =\frac{1}{1-\lambda}\ln \int_{\Lambda} [\rho(x)]^\lambda dx, \quad \lambda \ne 1
   \end{equation}
(whose limiting value $\lambda\rightarrow 1$ yields the Shannon entropy), and the Fisher information \cite{frieden_04,fisher} 
\begin{equation}
      F[\rho] =\int_{\Lambda} \frac{1}{\rho(x)}\left(\frac{d}{dx} \rho(x)\right)^2\,dx,
\end{equation}
which due to the involved derivative is a bit less global quantity.
Opposite to the variance, these information-theoretic spreading measures do not depend on any particular point of the interval $\Lambda$ being the domain of $\rho$. 
Note that all these measures
($V[\rho]$, $S[\rho]$, $R_{\lambda}[\rho]$, $F[\rho]$) are complementary, since each of them grasps a single different facet of the
probability density $\rho(x)$. Indeed, the variance measures the concentration of the density around the centroid while the R\'enyi and
Shannon entropies are measures of the extent to which the density is in fact concentrated. The Fisher informations is a quantitative estimation of the 
oscillatory character of the density since it measures the pointwise concentration of the probability over its support interval $\Lambda$.
\\

All measures of complexity mentioned in the Introduction (Cr\'amer-Rao, Fisher-Shannon and LMC) are defined as the products of two of the previously listed spreading measures. Each of them thus estimates the combined balance of two different facets of the probability density. The Crámer-Rao complexity \cite{dembo,dehesa_1,antolin_ijqc09}, which is defined as
\begin{equation}
     \label{Cramer_rao}
     C_{CR}\left[\rho\right]=F\left[\rho\right] \, V\left[\rho\right],
  \end{equation}
quantifies the gradient content of $\rho(x)$ jointly with the probability concentration around the centroid.  
The Fisher-Shannon complexity \cite{VignFS,angulo_pla08,romera_1}, which is given by  
  \begin{equation}
     \label{fishershannon}
      C_{FS}[\rho]=F[\rho] \times \frac{1}{2 \pi e} e^{2 S[\rho]},
  \end{equation}
measures the gradient content of $\rho(x)$ together with its total extent in the support interval.  
Finally, the biparametric LMC complexity (or LMC-Rényi complexity) \cite{pipek,lopez,lopezr,romera08} is:
\begin{equation}
     \label{lmcrenyi}
C_{\alpha,\beta}[\rho]
= e^{R_\alpha[\rho]-R_\beta[\rho]}, \quad 0 < \alpha < \beta, \quad \alpha,\beta\neq 1.
\end{equation}
Note that the case ($\alpha\rightarrow 1$, $\beta =2$) corresponds to the plain LMC complexity measure \cite{catalan_pre02} $C_{1,2}[\rho] = D[\rho] \times e^{S[\rho]}$, which measures the combined balance of the average height of $\rho(x)$ (also called disequilibrium $D[\rho] = e^{-R_2[\rho]}$), and its total extent.\\

These three complexity measures are known to be (a) dimensionless, (b) bounded from below by unity \cite{dembo,guerrero}, and (c) invariant
under translation and scaling transformation \cite{yamano_jmp04,yamano_pa04}. Moreover, the question whether the complexity measures are minimum for the two extreme (or \textit{least complex}) distributions corresponding to perfect order and maximum disorder (associated to a extremely
localized Dirac delta distribution and a highly flat distribution in the one dimensional case, respectively) is a long standing and controverted issue, not yet solved (see e.g. \cite{sanchez-moreno}). 

\section{Monotonicity of the complexity measures}
%
%
%
%
In this Section we investigate whether the complexity measures of Cr\'amer-Rao, Fisher-Shannon, and LMC-Rényi types given by the expressions (\ref{Cramer_rao}), (\ref{fishershannon}) and (\ref{lmcrenyi}), respectively, are complexity monotones.

\subsection{Fisher-Shannon complexity}

Let us first prove that for $\rho(x)$ with unbounded support, the Fisher-Shannon complexity $C_{FS}[\rho]$ given by (\ref{fishershannon}) is monotonic in the previously specified sense, with the family $\Xi$ of non-complex states formed by all the Gaussian densities (with arbitrary mean value and variance). The relevant operations $\mathcal{G}$ preserving  $\Xi$ are in this case constructed in terms of the convolution of a given distribution with some (once more arbitrary) Gaussian density. As the convolution of two Gaussians is another Gaussian, the family $\Xi$ is properly preserved.

The required monotonicity means that


\[
C_{FS}[\rho_\tau] \le C_{FS}[\rho],
\]
where $\rho_\tau=\mathcal{G}[\rho]$ is the convolution of $\rho$ with a Gaussian of variance $\tau$ (the mean value does not play any role). Taking into account the known properties of the Gaussian densities (convergence to the Dirac delta distribution) and the convolution, we have that
\[
\rho = \lim_{\tau\to 0}\rho_\tau,
\]
so that it is sufficient to show that $C_{FS}[\rho_\tau]$ is a decreasing function of $\tau$; that is,
\begin{equation}
\frac{d}{d\tau} C_{FS}[\rho_\tau]\le 0.
\label{eq:cfs_decreasing}
\end{equation}
To achieve that goal, we recall the de Bruijn identity \cite{cover}
\begin{equation}
\frac{d}{d\tau} S[\rho_\tau] = \frac12 F[\rho_\tau],
\label{eq:debruijn}
\end{equation}
which implies that
\[
C_{FS}[\rho_\tau] = \frac{d}{d\tau} N[\rho_\tau],
\]
where $N[\rho] \equiv \frac{e^{2S[\rho]}}{2\pi e}$ denotes the entropy power of $\rho$. The desired inequality (\ref{eq:cfs_decreasing}) follows from the concavity of the entropy power
\[
\frac{d^2}{d\tau^2} N[\rho_\tau] \le 0,
\]
which was proved by Costa \cite{costa}.

\subsection{Crámer-Rao complexity}
Studying the second quantity, namely
the Crámer-Rao complexity given by (\ref{Cramer_rao}), we show that it is monotonic in the same sense as the Fisher-Shannon complexity measure.
Analogously to the Fisher-Shannon case, we want to prove that
\begin{equation}
\frac{d}{d\tau} C_{CR}[\rho_\tau] \le 0.
\label{eq:cr_decreasing}
\end{equation}

Using again the de Bruijn identity (\ref{eq:debruijn}) together with the known relation
\[
V[\rho_\tau] = V[\rho] + \tau,
\]
we obtain
\[
\frac{d}{d\tau}C_{CR}[\rho_\tau]
=
2\frac{d}{d\tau}S[\rho_\tau]
+2(V[\rho]+\tau)
\frac{d^2}{d\tau^2}S[\rho_\tau].
\]
The concavity of the entropy power implies that:
\[
\frac{d^2}{d\tau^2}N[\rho_\tau]=
2N[\rho_\tau]\left[
2\left(\frac{d}{d\tau}S[\rho_\tau]\right)^2
+\frac{d^2}{d\tau^2}S[\rho_\tau]
\right]\le 0,
\]
from which we get the inequality
\[
\frac{d^2}{d\tau^2}S[\rho_\tau]\le - 2\left(\frac{d}{d\tau}S[\rho_\tau]\right)^2.
\]
This inequality together with the de Bruijn identity (\ref{eq:debruijn}) yield the relation
\begin{equation}
\frac{d}{d\tau}C_{CR}[\rho_\tau]
\le
F[\rho_\tau]-(V[\rho]+\tau)(F[\rho_\tau])^2.
\label{eq:inequality_cr}
\end{equation}

The right-hand side of (\ref{eq:inequality_cr}) is a negative function of $F[\rho_\tau]$ provided that
\[
F[\rho_\tau]\ge \frac{1}{2(V[\rho]+\tau)}.
\]
But the last inequality is always satisfied due to the Crámer-Rao bound \cite{dembo}
\begin{equation}
F[\rho_\tau]\ge   \frac{1}{V[\rho_\tau]} = \frac{1}{V[\rho]+\tau}.
\label{eq:cramerrao_inequality}
\end{equation}
Moreover, if the Crámer-Rao bound becomes saturated, the right-hand side of (\ref{eq:inequality_cr}) vanishes. 
Since the derivative of $C_{CR}[\rho_\tau]$ is upper-bounded by a decreasing function of $F[\rho_\tau]$ whose maximum value is equal to zero, the proposed inequality (\ref{eq:cr_decreasing}) is proved.

\subsection{LMC-Rényi complexity}
In the last part we shall discuss 
the generalized LMC complexity measure $C_{\alpha,\beta}[\rho]$. 
This measure becomes relevant when the set $\Xi$ of non-complex states is assumed to contain all the uniform densities $\rho(x)=1/L$ with bounded (not necessarily compact) support of length $L$. In the limit of a very narrow, compact support ($L \rightarrow 0$) the non-complex density becomes the Dirac delta distribution concentrated in the center of the support.

A more difficult question is about the complete family of operations $\mathcal{G}$
preserving the class $\Xi$. In order to better understand the non-complex
landscape described by $\Xi$, we will study the simplest, discrete
counterpart scenario in dimension two. In this case [instead of $\rho\left(x\right)$]
one uses a collection of two probabilities $\left(p,1-p\right)$,
given in terms of a single number $0\leq p\leq1$. In this simplified
situation there are three non-complex states given by the values $p\in\left\{ 0,1/2,1\right\}$.

We start by an assumption that any allowed operation performed on
discrete probability distributions can be represented by a stochastic
matrix. Thus, in the two-dimensional case every operation applicable
to the probability vector is of the form
\begin{equation}
\left(\begin{array}{cc}
a & b\\
1-a & 1-b
\end{array}\right),
\end{equation}
with $0\leq a,b\leq1$. There are only four matrices, different from
the identity, which preserve the set of the three non-complex probability
vectors:
\begin{equation}
\left(\begin{array}{cc}
0 & 0\\
1 & 1
\end{array}\right),\quad\left(\begin{array}{cc}
1 & 1\\
0 & 0
\end{array}\right),\quad\frac{1}{2}\left(\begin{array}{cc}
1 & 1\\
1 & 1
\end{array}\right),\quad\left(\begin{array}{cc}
0 & 1\\
1 & 0
\end{array}\right).
\end{equation}
The first three operations simply output each of the non-complex vectors, independently of the input. The last one is the permutation.

While increasing the dimension (but staying on the discrete ground) the number of allowed operations also increases, but the qualitative description of the set $\mathcal{G}$ remains the same. In order to preserve the class of non-complex states one can either transform any probability vector to become the member of the class $\Xi$, or freely permute the probability components. At the end of the day, we are left with a conclusion that only the permutations provide non-trivial but allowed transformations of the set $\Xi$.

But the discrete counterpart $C_{\alpha,\beta}[q]$ of the generalized LMC complexity measure (given by discrete R\'enyi entropies of the probability vector $q$) is invariant with respect to permutations. This fact implies that in the discrete scenario the LMC measure somehow trivially satisfies the monotonicity requirement. For any probability vector $q$ either  $C_{\alpha,\beta}[\mathcal{G}[q]]=C_{\alpha,\beta}[q]$ or $C_{\alpha,\beta}[\mathcal{G}[q]]=1$.

The remaining question is if the discrete analysis discussed above can directly be transferred to the continuous scenario. It seems to be very likely that the measure $C_{\alpha,\beta}[\rho]$ is a monotone in the same sense as $C_{\alpha,\beta}[q]$. On the other hand, the fundamental entropic uncertainty relation \cite{BBM} shows that the continuous limit does not need to be direct, since between the uniform and the peaked distribution there is room for the Gaussians. We shall thus leave the construction of the full class $\mathcal{G}$ in the continuous case, as well as the rigorous proof of the monotonicity for LMC, as two open questions for future research.

\section{Conclusions}

The purpose of this communication is to contribute to quantify how simple or how complex are the many-particle systems in terms of the one-particle probability density which, according to the density functional theory, characterize their physical and chemical properties. Since there does not exist a unique notion of complexity able to grasp our intuition in the appropriate manner it is important to better understand the conditions under which a given quantity is a proper measure of complexity. It is even more important as perhaps the most universal and appropriate descriptions of statistical complexity are not yet known.\\

To contribute to settle down this issue, we have introduced the mathematical notion of monotonicity of the complexity measure of a probability distribution patterned after the resource theory of quantum entanglement and coherence. We have discussed under what conditions the basic complexity measures of physical systems satisfy this requirement. As the main aim of the letter was to provide a general framework, the studies of particular examples, with special emphasis on the LMC complexity measure, shall be continued in the future. The mathematical results collected during the investigation of Gaussian-based monotonicity might be useful for other research such as studies of pointer-based measurements \cite{Frey}.

\section*{Acknowledgments}
 This work was partially supported by the Projects
P11-FQM-7276 and FQM-207 of the Junta de Andalucia, and by the MINECO grants FIS2011-24540, FIS2014-54497P and FIS2014-59311-P.
\L .R. acknowledges financial support by the grant number 2014/13/D/ST2/01886
of the National Science Center, Poland. Research in Freiburg is supported
by the Excellence Initiative of the German Federal and State Governments
(Grant ZUK 43), the Research Innovation Fund of the University of
Freiburg, the ARO under contracts W911NF-14-1-0098 and W911NF-14-1-0133
(Quantum Characterization, Verification, and Validation), and the
DFG (GR 4334/1-1). The work of I. V. Toranzo acknowledges partial support of the program FPU of MINECO.

\end{document}